\documentclass[twocolumn,showpacs,preprintnumbers,amsmath,amssymb]{revtex4-1}
\usepackage{graphicx}
\usepackage[sort&compress]{natbib}
\usepackage{subfigure}
\usepackage{amsmath}
\usepackage{amsfonts}
\usepackage{cancel}

\begin{document}
\preprint{APS/123-QED}
\arraycolsep1.5pt

\newcommand{\Ima}{\textrm{Im}}
\newcommand{\Rea}{\textrm{Re}}
\newcommand{\mev}{\textrm{ MeV}}
\newcommand{\be}{\begin{equation}}
\newcommand{\ee}{\end{equation}}
\newcommand{\ba}{\begin{eqnarray}}
\newcommand{\ea}{\end{eqnarray}}
\newcommand{\gev}{\textrm{ GeV}}
\newcommand{\nn}{{\nonumber}}
\newcommand{\dtres}{d^{\hspace{0.1mm} 3}\hspace{-0.5mm}}

\title{Determination of the real part of the $\eta^\prime$-Nb optical potential }
\author{
M.~Nanova$^{1}$,~S.~Friedrich$^{1}$,~V.~Metag$^{1}$,~E.~Ya.~Paryev$^{2}$,~F.~N.~Afzal$^{3}$,~D.~Bayadilov$^{3,4}$,~R.~Beck$^{3}$,~M.~Becker$^{3}$,~S.~B\"ose$^{3}$,\\K.-T.~Brinkmann$^{1}$,~V.~Crede$^{5}$,~D.~Elsner$^{6}$,~F.~Frommberger$^{6}$,~M.~Gr\"uner$^{3}$,~E.~Gutz$^{1}$,~Ch.~Hammann$^{3}$,~J.~Hannappel$^{6}$,\\J.~Hartmann$^{3}$,~W.~Hillert$^{6}$,~P.~Hoffmeister$^{3}$,~Ch.~Honisch$^{3}$,~T.~Jude$^{6}$,~F.~Kalischewski$^{3}$,~I.~Keshelashvili$^{7,a}$,~F.~Klein$^{6}$,\\K.~Koop$^{3}$,~B.~Krusche$^{7}$,~M.~Lang$^{3}$,~K.~Makonyi$^{1,b}$,~F.~Messi$^{6}$,~J.~M\"uller$^{3}$,~J.~M\"ullers$^{3}$,~D.-M.~Piontek$^{3}$,~T.~Rostomyan$^{7}$,\\D.~Schaab$^{3}$,~C.~Schmidt$^{7}$,~H.~Schmieden$^{6}$,~R.~Schmitz$^{3}$,~T.~Seifen$^{3}$,~C.~Sowa$^{8}$,~K.~Spieker$^{3}$,~A.~Thiel$^{3}$,~U.~Thoma$^{3}$,\\T.~Triffterer$^{8}$,~M.~Urban$^{3}$,~H.~van~Pee$^{3}$,~D.~Walther$^{3}$,~C.~Wendel$^{3}$,~D.~Werthm\"uller$^{7,c}$,~U.~Wiedner$^{8}$,~A.~Wilson$^{3}$,\\L.~Witthauer$^{7}$,~Y.~Wunderlich$^{3}$, and H.-G.~Zaunick$^{1}$
(The CBELSA/TAPS Collaboration)}

\affiliation{
{$^{1}$II. Physikalisches Institut, Universit\"at Gie{\ss}en, Germany}\\
{$^{2}$ Institut of Nuclear Research, Russian Academy of Sciences, Moscow, Russia}\\
{$^{3}$Helmholtz-Institut f\"ur Strahlen- und Kernphysik, Universit\"at Bonn, Germany }\\
{$^{4}$Petersburg Nuclear Physics Institute, Gatchina, Russia} \\
{$^{5}$Department of Physics, Florida State University, Tallahassee, FL, USA}\\
{$^{6}$Physikalisches Institut, Universit\"at Bonn, Germany}\\
{$^{7}$Departement Physik, Universit\"at Basel, Switzerland}\\
{$^{8}$Physikalisches Institut, Universit\"at Bochum, Germany}\\
{$^{a}$Current address: Institut f\"ur Kernphysik, Forschungszentrum J\"ulich}, Germany\\
{$^{b}$Current address: Stockholm University, Stockholm, Sweden}\\
{$^{c}$Current address: School of Physics and Astronomy, University of Glasgow, UK}\\
}

\date{\today}
%
\begin{abstract}
The excitation function and momentum distribution of $\eta^\prime$ mesons have been measured in photoproduction off $^{93}$Nb in the energy range of 1.2-2.9 GeV. The experiment has been performed with the combined Crystal Barrel and MiniTAPS detector system, using tagged photon beams from the ELSA electron accelerator. Information on the sign and magnitude of the real part of the $\eta^\prime$-Nb potential has been extracted from a comparison of the data with model calculations. An attractive potential of -($41 \pm$10(stat)$\pm$15(syst)) MeV depth at normal nuclear matter density is deduced within model uncertainties. This value is consistent with the potential depth of -($37 \pm $10(stat)$\pm$10(syst)) MeV obtained in an earlier measurement for a light nucleus (carbon). This relatively shallow $\eta^\prime$-nucleus potential will make the search for $\eta^\prime$ - nucleus bound states more difficult.
 \end{abstract}
\maketitle

\section{Introduction}
\label{Intro}
The masses and the excitation spectrum of baryons and mesons are an important testing ground for our understanding of the dynamics of quarks and gluons in the non-perturbative regime of Quantum-Chromodynamics. A profound understanding of the excitation energy spectrum of baryons composed of up, down, and strange quarks is still lacking. It remains a challenge to link the empirical excitation spectrum to theoretical predictions and to unravel the relevant degrees of freedom. Although the constituent quark model has many successes, detailed studies of nucleon excitations have provided evidence that some of the low-lying excited states may have a structure which goes beyond the simple 3-quark configuration. Kaiser et al. \cite{Kaiser} discuss the possibility that for example the S$_{11}$(1535) resonance may be a dynamically generated quasi-bound K$\Sigma$-K$\Lambda$ state. Similar interpretations have been proposed for the $\Lambda$(1405) resonance as having a $\bar{K}$-N and $\pi$-$\Sigma$ molecular structure \cite{Oset,Meissner,Thomas}. The recently observed resonances consistent with pentaquark states P$_C$(4380) and P$_C$(4450) \cite{LHCb} may also be dynamically generated baryon-meson molecular configurations \cite{Wu}. Hadronic degrees of freedom like the meson-baryon interaction may thus play an important role in the structure of excited N and  $\Lambda$ states. 

The meson-baryon interaction has been investigated experimentally in near-threshold meson production to determine the meson-nucleon scattering length which is a measure for the strength of the interaction. Scattering lengths for the $\eta$-N \cite{Arndt}, K-N \cite{Iwasaki}, $\omega$-N \cite{Strakovsky}, and $\eta^\prime$-N  \cite{Moskal} systems have been deduced. 

If bound or quasi-bound meson-nucleon configurations exist a next step would be to ask whether also bound systems of mesons and nucleon clusters may exist. The possible existence of compact K$^-$pp clusters was proposed by Yamazaki and Akaishi \cite{Yamazaki_Akaishi} and has attracted a lot of attention experimentally and theoretically. Following first claims of observing kaonic clusters \cite{FINUDA,DISTO} conflicting results have been reported and the existence of such states discussed controversially (see recent publications \cite{Ichikawa,HADES,Fabbietti} and references cited therein). The binding energy of these states may not be very large and they may have a rather large width which makes it difficult to detect them experimentally. 

Another step further is the quest for the possible existence of meson-nucleus bound states. Deeply bound pionic states have been observed \cite{Itahashi, Geissel}. These are halo-like configurations with a $\pi^-$ meson bound in a potential pocket at the nuclear surface generated by the superposition of the attractive Coulomb interaction and the repulsive s-wave $\pi^-$-nucleus interaction \cite{Kienle_Yamazaki}. 

We are interested in the question whether the strong interaction alone is strong enough to form meson-nucleus bound states. This can be tested by looking for bound states of neutral mesons with nuclei. In order to find out which neutral meson is the most promising candidate for observing mesic states the meson-nucleus interaction has to be studied. This interaction can be described by an optical potential  \cite{Nagahiro1}
\begin{equation}
U(r) = V(r) + iW(r),
\end{equation}
where $V$ and $W$ denote the real and imaginary parts of the optical potential, respectively, and $r$ is the distance between the meson and the centre of the nucleus.
\par
The strength of the real part of the meson-nucleus potential is connected to the meson in-medium mass shift $\Delta m(\rho_{0})$ at saturation density $\rho_{0}$ \cite{Nagahiro1}
\begin{equation}
V(r) = \Delta m(\rho_{0})\cdot c^2\cdot \frac{\rho(r)}{\rho_{0}}.
\end{equation}
The imaginary part of the potential describes the meson absorption in the medium via inelastic channels and is related to the in-medium width $\Gamma_{0}$ of the meson at nuclear saturation density by \cite{Nanova_tr}
\begin{equation}
W(r) = -\frac{1}{2}\Gamma_{0}\cdot \frac{\rho(r)}{\rho_{0}}.
\end{equation}

In recent photoproduction experiments, we studied the $\omega$- and $\eta^\prime$- nucleus interaction and deduced information on the real \cite{Nanova_realC,Friedrich,Metag_PPNP,Metag_HypInt} and imaginary part \cite{Nanova_tr,Kotulla,Kotulla_err} of the optical potential. For the latter, values of $\approx$ -70 MeV and $\approx$ -10 MeV were extracted for the $\omega$ and $\eta^\prime$ meson, respectively, at saturation density and for average recoil momenta of $\approx$ 1~GeV/c from transparency ratio measurements by studying the attenuation of the meson flux in the photoproduction off various nuclei.  The real part of the meson-nucleus optical potential has, however, so far only been determined for a light nucleus (carbon). In the present work we extend these studies for the $\eta^\prime$ meson to a heavier nucleus (Nb, A=93) to investigate whether there is any dependence of the optical model parameters on the nuclear mass number A.

The paper is structured as follows: The experimental set up and the conditions of the experiment are described in section II. Details of the analysis are given in section III. In section IV the results are presented and compared to theoretical calculations in section V. Concluding remarks are given in section VI.

\section{Experiment}
\label{sec:exp}
The experiment was performed at the ELSA electron accelerator facility~\cite{Husmann_Schwille,Hillert} at the University of Bonn. Tagged photons of energies 1.2-2.9~GeV were produced via bremsstrahlung from an electron beam of 3.0 GeV, scattered off a diamond radiator (500 $\mu$m thick). The energy of generated photons was determined by a tagging hodoscope with an energy resolution better than 0.4$\% \cdot$E$_{\gamma}$. The photon beam, collimated by an aperture of 7 mm diameter,  impinged on a 1 mm thick $^{93}$Nb target (8.6$\%$ of a radiation length $X_0$). Decay photons from mesons produced in the target were registered in the combined Crystal Barrel (CB) and MiniTAPS  detector system. The CB detector \cite{Aker},  a homogeneous electromagnetic calorimeter, consisted of 1230 CsI(Tl) crystals read out with photodiodes, subtending polar angles of 29$^\circ$-156$^\circ$. In the forward angular range between 11$^\circ$ and 28$^\circ$,~90 CsI(Tl) crystals (Forward Plug (FP)) were mounted and read out with photomultipliers (PMT) providing energy and time information. The angular range between 1$^\circ$ and 11$^\circ$  was covered by the MiniTAPS forward wall  \cite{Novotny,Gabler} at a distance of 210 cm from the centre of the CB, consisting of 216 BaF$_2$ crystals, read out via PMTs with electronics described in~\cite{Drexler}. The high granularity and the large solid angle coverage made the detector system ideally suited for the detection and reconstruction of multi-photon events.

For charged particle identification each BaF$_2$ module of the MiniTAPS array and the 90 CsI(Tl) crystals of the FP were equipped with plastic scintillators. In the angular range of 23$^\circ$-167$^\circ$ a three layer fibre detector with 513 scintillating fibres, surrounding the target and placed at the centre of the CB, served for charged particle detection~\cite{Suft}. To suppress electromagnetic background at forward angles, a gas-Cherenkov detector with an index of refraction of n=1.00043 was mounted in front of the MiniTAPS array. 

In order to improve the statistics at low $\eta^\prime$ momenta, the orientation of the diamond radiator was chosen to generate an excess of coherent photons peaking at an energy of 1.5 GeV in addition to the 1/E$_{\gamma}$  bremsstrahlung flux distribution. The polarisation of the radiation was not exploited in the analysis of the data. The photon flux through the target was determined by counting the photons reaching the gamma intensity monitor (GIM) at the end of the setup in coincidence with electrons registered in the tagging system. The total rate in the tagging system was $\approx$10 MHz. The dead time introduced by the gas-Cherenkov detector was about 25$\%$. The GIM dead time, corrected for in the flux determination, was about 20$\%$. 

Online event selection was made using first- and second-level triggers. The detectors contributing to the first-level trigger were the FP, MiniTAPS and gas-Cherenkov together with signals from the tagger. CB could not be used in the first-level trigger because of the long rise time of the photodiode signals. The second-level trigger was based on a FAst Cluster Encoder (FACE), providing the number of clusters in the CB within $\approx $ 10~$\mu$s. Events with at least two hits in the calorimeters and no hit in the gas-Cherenkov detector were selected for further processing. The events were collected in a data taking period of  960~h.  More details on the experimental setup and the running conditions can be found in \cite{Nanova_realC,Thiel}.

\section{Data Analysis}
\label{sec:ana}
\begin{figure}
 \resizebox{0.4\textwidth}{!}
  {
   \includegraphics[height=0.7\textheight]{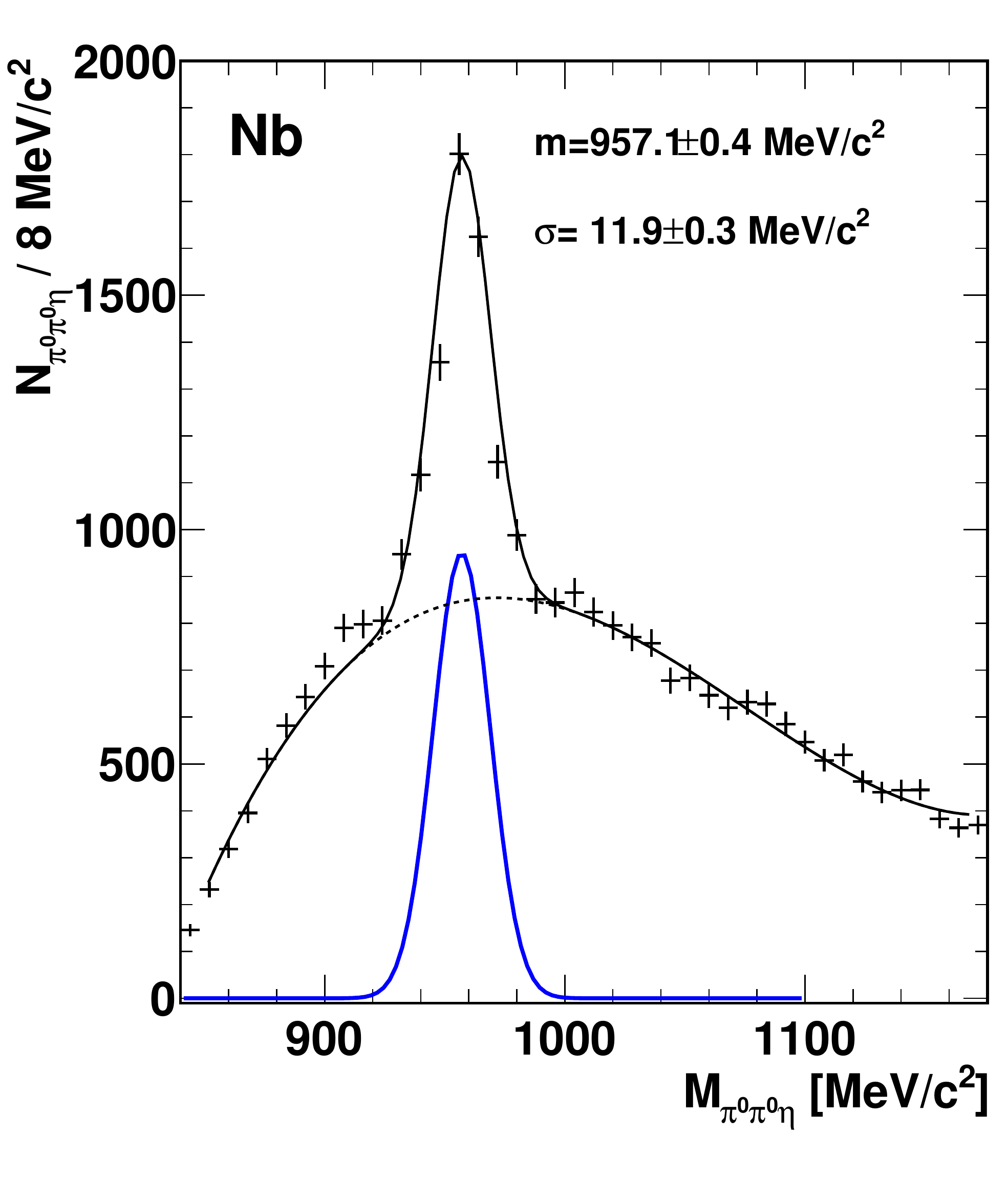}
  }
\caption{(Color online) The $\pi^0\pi^0\eta$ invariant mass distribution measured in photoproduction off Nb in the incident photon energy range of 1.2-2.9 GeV. The solid curve represents a fit to the data using a Gaussian function combined with a polynomial function for the background. The fit parameters are: $\sigma$=11.9$\pm$0.3 MeV/$c^2$ (corresponding to the instrumental resolution), m=957.1$\pm$0.4 MeV/$c^2$.}
\label{fig:etaprime_signal}
\end{figure}

The $\eta^\prime$ mesons were identified via their $\eta^\prime\rightarrow \pi^{0}\pi^{0}\eta \rightarrow 6 \gamma$ decay with an overall branching ratio of 8.5$\%$ \cite{PDG}. For the reconstruction of $\eta^\prime$ mesons from the registered decay photons, only events with 6 neutral and any number of charged hits and with an energy sum of neutral clusters larger than 600 MeV were used. The 6 photons were combined in two pairs of two photons with invariant masses in the range 115 MeV/$c^2 \le m_{\gamma\gamma} \le$ 155 MeV/$c^2$ (corresponding to a $\pm 3 \sigma$ cut around $m_{\pi^{0}}$) and one pair with invariant mass in the range 510 MeV/$c^2 \le m_{\gamma\gamma} \le$ 590 MeV/$c^2$ (roughly corresponding to a $\pm 2 \sigma$ cut around $m_{\eta}$). The best photon combination was selected based on a $\chi^2$ minimisation. To suppress the background from $\eta \rightarrow 3\pi^{0}$ decays, events with three photon pairs with an invariant mass close to the pion mass ($m_{\pi^{0}}$) were removed from the data set. Random coincidences between the tagger and the detector modules in the first level trigger were removed by a cut in the corresponding time spectra. The resulting $\pi^{0}\pi^{0}\eta$ invariant mass spectrum is shown in Fig.~\ref{fig:etaprime_signal}. 
The $\pi^{0}\pi^{0}\eta$ invariant mass spectrum was fitted with a Gaussian function and a polynomial to describe the background. The $\eta^\prime$-signal in the $\pi^{0}\pi^{0}\eta$ spectrum had a width $\sigma$=$11.9\pm0.3$ MeV/$c^2$ and a position $m=957.1\pm0.4$ MeV/$c^2$, in good agreement with the PDG value \cite{PDG}. 
In total, $\approx$ 3500 $\eta^\prime$ mesons were reconstructed in the photon energy range 1.2-2.9~GeV. 

\begin{figure}
 \resizebox{0.5\textwidth}{!}
  {
   \includegraphics[height=0.9\textheight]{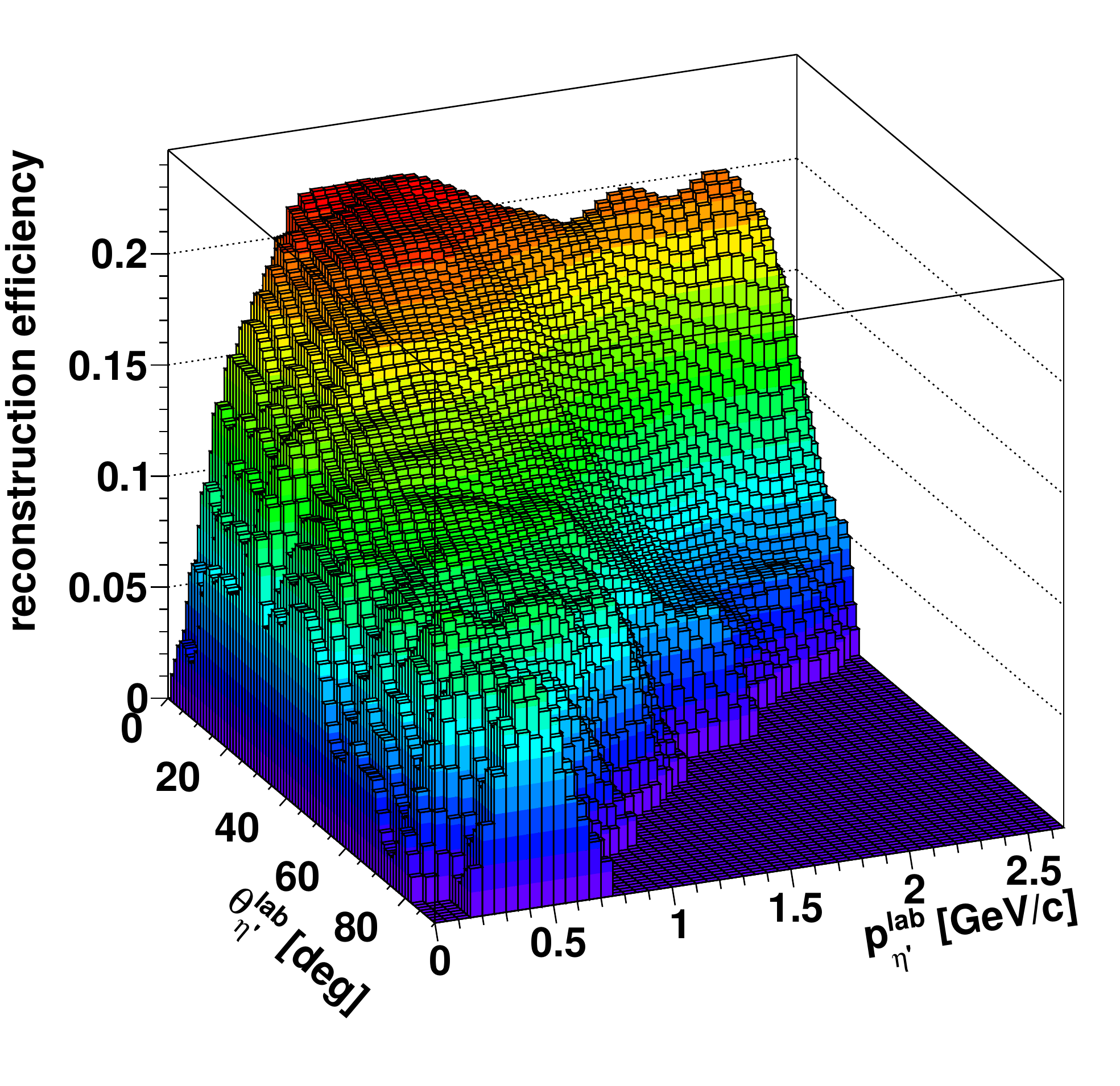}
  }
\caption{(Color online) Two-dimensional reconstruction efficiency for $\eta^\prime$ photoproduction off Nb as a function of the $\eta^\prime$ momentum and laboratory angle for the incident photon energy range of 1.2-2.9 GeV.}
\label{fig:etaprime_acc}
\end{figure}

For the determination of the angle differential and total cross sections, the efficiency for reconstructing the reaction of interest has to be known. Applying the GEANT3 package \cite{GEANT} with a full implementation of the detector system, the reaction $\gamma \text{Nb} \rightarrow \text{X}\eta^\prime$ was simulated, using as input the measured angular distributions of $\eta^\prime$ mesons produced off protons and neutrons bound in deuterium \cite{Igal}. In addition, the Fermi motion of nucleons in the target nucleus, as parameterised by \cite{Cioffi}, has been taken into account. The reconstruction efficiency was determined as a function of the laboratory angle and the momentum of the $\eta^\prime$ meson. This approach ensured that the appropriate acceptance was used even if the angle and momentum of the $\eta^\prime$ meson deviated from the kinematics of the reaction because of final state interactions (FSI) in the nuclear environment. For the $\eta^\prime$ meson, FSI effects are, however, expected to be small because of the rather small cross section for elastic $\eta^\prime$ scattering predicted to be $\sigma_{\text{el}}^{\eta^\prime} \approx 3$~mb \cite{Oset_Ramos}. This corresponds to a mean free path of $\approx$ 20~fm which is large compared to nuclear dimensions. The reconstruction efficiency was determined by taking the ratio of the number of reconstructed and the number of generated $\gamma \text{Nb} \rightarrow \eta^\prime \text{X}$ events in the $\eta^\prime\rightarrow \pi^0\pi^0\eta \rightarrow 6 \gamma$ channel for each angular- and momentum bin. The resulting reconstruction efficiency $\epsilon_{\gamma \text{Nb} \rightarrow \eta^\prime \text{X}} (p_{\eta^\prime}^{\text{lab}},\theta _{\eta^\prime}^{\text{lab}})$ varies smoothly over the full kinematic range as shown in Fig.~\ref{fig:etaprime_acc}, for the incident photon energy range of 1.2-2.9~GeV. In the Monte Carlo simulations, the same trigger conditions as in the experiment were applied. \\

For the cross section determinations, the $\pi^0 \pi^0 \eta$ invariant mass histograms were filled with an event-by-event weighting by the inverse photon flux N$_{\gamma}$ and the reconstruction efficiency $\epsilon_{\gamma \text{Nb} \rightarrow \eta^\prime \text{X}} (p_{\eta^\prime}^{\text{lab}},\theta _{\eta^\prime}^{\text{lab}})$ for each bin in $\eta^\prime$ momentum $p_{\eta^\prime}^{\text{lab}} $ and angle $\theta _{\eta^\prime}^{\text{lab}}$ in the laboratory frame. The differential $\eta^\prime$ cross sections were determined by applying the same fit procedure for 8 bins of the incident photon energy and for 5 bins of $\cos\theta^{\text{c.m.}}_{\eta^\prime}$, where $\theta^{\text{c.m.}}_{\eta^\prime}$ is the angle of the $\eta^\prime$ in the centre of mass system of the incident photon and a target nucleon at rest, neglecting Fermi motion. The polar angular binning was chosen according to the available statistics and was larger than the angular resolution in the c.m. system. The statistical errors were determined from the yield of the $\eta^\prime$ signal (S) in each energy and $\cos\theta^{\text{c.m.}}_{\eta^\prime}$ bin and the counts in the background below the peak (BG) according to the formula: $\Delta \text{N} = \sqrt{(\text{S+BG})}$. The total cross section for $\eta^\prime$ photoproduction was determined (i) by integrating the differential cross sections and (ii) by direct determination of the $\eta^\prime$ meson yields for different incident photon energy bins. The two methods were applied as a systematic check of the fit procedure to extract the $\eta^\prime$ invariant mass signal over different kinematic ranges. The results are compared and further discussed in sec. \ref{sec:tot}.
\par 

The different sources of systematic errors are summarised in Table \ref{tab:syst}.  The systematic errors in the fit procedure were estimated to be in the range of 10-15\% by applying different background functions and fit intervals. Varying the start distributions in the acceptance simulation between isotropic and forward peaking $\eta^\prime$ angular distributions, the systematic errors of the acceptance determination were determined to be less than 10\%. The photon flux through the target was measured by counting the photons reaching the GIM in coincidence with electrons registered in the tagger system. Systematic errors in the photon flux determination after dead time correction were estimated to be about 5-10\%. The systematic errors introduced by uncertainties in the photon shadowing  (see below) were $\approx$ 10\%. The total systematic error of  the cross section determinations, obtained by adding the systematic errors quadratically, was 23\%.

\begin{table}[h!]
\centering
\caption{Sources of systematic errors}
\begin{footnotesize}

\begin{tabular}{cc}
\hline
fits & $\approx$ 10-15\%\\
reconstruction efficiency & $\lesssim$ 10\%\\
photon flux & 5-10\%\\
photon shadowing & $\approx$ 10\%\\
\hline
total& $\approx$ 23\%\\
\end{tabular}
\end{footnotesize}

\label{tab:syst}
\end{table}

\section{Experimental results}
\subsection{Differential cross sections for the $\eta^\prime$ photoproduction off Nb}
\label{sec:diff}
The differential cross sections have been determined according to:
\begin{eqnarray}
\frac{d\sigma}{d (\cos \theta_{\eta^\prime}^{\text{c.m.}})} = \sum_{\text{p}_{\eta^\prime}^{\text{lab}}} \frac{\text{N}_{\eta^\prime\rightarrow \pi^0 \pi^0 \eta}(\text{p}_{\eta^\prime}^{\text{lab}}, \theta_{\eta^\prime}^{\text{lab}})}{ \epsilon_{\gamma \text{Nb}\rightarrow \eta^\prime \text{X}} (\text{p}_{\eta^\prime}^{\text{lab}}, \theta_{\eta^\prime}^{\text{lab}})} \nonumber \\
\cdot \frac{1}{\text{N}_{\gamma} \cdot \text{n}_{\text{t}}} \cdot \frac{1}{\Delta \cos\theta_{\eta^\prime}^{\text{c.m.}}} \cdot \frac{1}{\frac{\Gamma_{\eta^\prime \rightarrow \pi^0 \pi^0 \eta \rightarrow 6 \gamma}}{\Gamma_{\text{total}}}},
\end{eqnarray}
where N$_{\eta^\prime \rightarrow \pi^0 \pi^0 \eta}(p_{\eta^\prime}^{\text{lab}}, \theta_{\eta^\prime}^{\text{lab}})$ is the number of reconstructed $\eta^\prime$ mesons extracted by the fit procedure as described in Sec.~\ref{sec:ana} in each (p$_{\eta^\prime}^{\text{lab}}, \theta_{\eta^\prime}^{\text{lab}}$) bin; N$_{\gamma}$ is the photon flux; n$_{\text{t}}$ is the density of the target nucleons multiplied by the target thickness (5.55$\cdot$10$^{21}$cm$^{-2}$); $\Delta \cos\theta_{\eta^\prime}^{\text{c.m.}}$ is the angular bin in the c.m. frame; $\frac{\Gamma_{\eta^\prime \rightarrow \pi^0 \pi^0 \eta \rightarrow 6 \gamma}}{\Gamma_{\text{total}}}$ is the decay branching fraction of 8.5\% for the decay channel $\eta^\prime \rightarrow \pi^0 \pi^0 \eta \rightarrow 6 \gamma$.\\

\par 
\begin{figure}
 \resizebox{0.5\textwidth}{!}
 {
   \includegraphics[height=0.9\textheight]{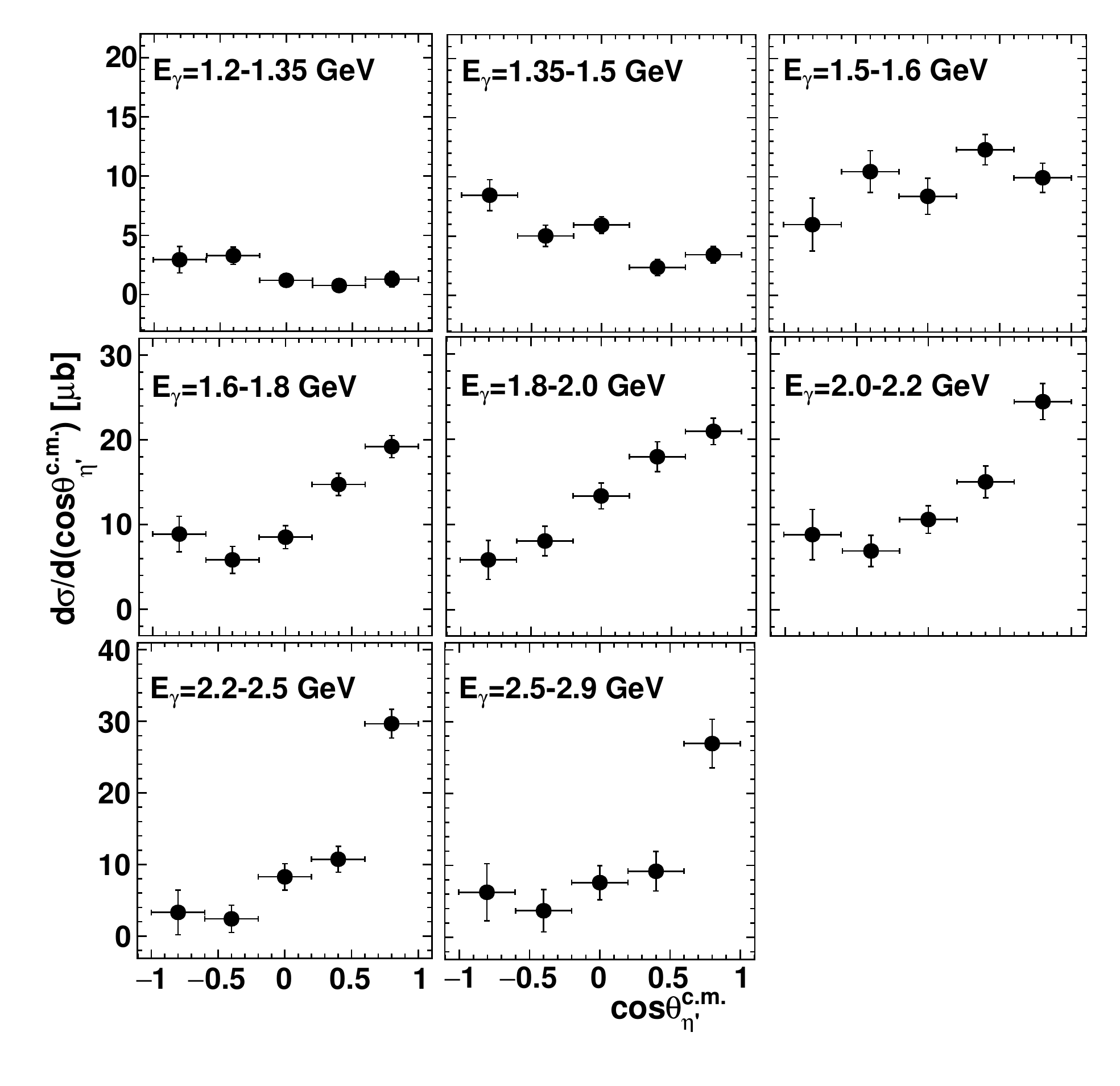} 
     }
\caption{Differential cross sections for photoproduction of $\eta^\prime$ mesons off Nb for different bins in the incident photon energy range 1.2-2.9 GeV determined in five $\cos\theta_{\eta'}^{\text{c.m.}}$ bins of width 0.4.}
\label{fig:diffcs}
\end{figure}

Fig.~\ref{fig:diffcs} presents the differential cross sections $d\sigma /d (\cos\theta^{\text{c.m.}}_{\eta^\prime})$ for 8 bins in the incident photon energy range. The dead time of the gas-Cherenkov detector and the GIM have been corrected for. Furthermore, the reduction in the incident photon flux due to photon shadowing has been taken into account by multiplying the observed $\eta^\prime$ yield by 1.17 \cite{Falter,bianchi,Muccifora}. A rather flat angular distribution is observed at low energies near the production threshold on a free nucleon ($E_{\gamma}^{\text{thr}}$=1.447~GeV). For higher photon energies E$_{\gamma} \ >$  1.8~GeV, the angular distributions show a forward rise, characteristic for t-channel production. This behaviour is similar to previous results on angular distributions for $\eta^\prime$ photoproduction off carbon \cite{Nanova_realC}.

\subsection{Total cross section for the $\eta^\prime$ photoproduction off Nb}
\label{sec:tot}

The total cross section for the $\eta^\prime$ photoproduction off Nb is shown in Fig.~\ref{fig:tot_Nb} (Left). The integration of the differential cross sections and the direct determination of the cross section from the $\eta^\prime$ yield in different incident photon energy bins give consistent results within errors. The cross section is found to be non-zero below E$_{\gamma}$ = 1.447~GeV, the threshold energy for photoproduction of $\eta^\prime$ mesons off the free nucleon. This is on the one hand, due to the Fermi motion of nucleons in the Nb target which gives rise to a distribution of the energy $\sqrt\text{s}$ available in the centre-of-mass system for a given incident photon energy. On the other hand, also the mass of the meson might drop in a nuclear medium - as discussed below - which lowers the production threshold and increases the phase space for meson production below the free threshold energy.

\section{Comparison to the theoretical model predictions and previous experimental results}
Weil et al. \cite{Weil} discussed the possibility to extract information on the in-medium meson mass and the real part of the meson-nucleus potential from a measurement of the excitation function and/or momentum distribution of mesons in the photoproduction off a nucleus. A lowering of the meson mass in the medium decreases the meson production threshold and the enlarged phase space will consequently increase the production cross section for a given incident beam energy as compared to a scenario without mass shift. The lowering of the meson mass in the medium also affects the momentum distribution of the produced meson in the final state. When a meson is produced with a lower mass, then its total energy is on average also reduced due to kinematics. In addition, mesons produced within the nuclear medium must regain their free mass upon leaving the nucleus. Thus, in case of an in-medium mass drop, this mass difference has to be compensated at the expense of their kinetic energy.  As demonstrated in GiBUU transport-model calculations~\cite{Weil}, this leads to a downward shift in the momentum distribution for near-threshold energies as compared to a scenario without mass shift. A mass shift can thus be indirectly inferred from a measurement of the excitation function as well as from the momentum distribution of the meson. This idea, initially worked out for $\omega$ mesons \cite{Weil}, has independently been pursued on a quantitative level for $\eta^\prime$ mesons by Paryev~\cite{Paryev}.

\subsection{Excitation function for  $\eta^\prime$ mesons}
\label{sec:excit}
\par
\begin{figure*}
\begin{center}
\resizebox{1.\textwidth}{!}
  {
   \includegraphics{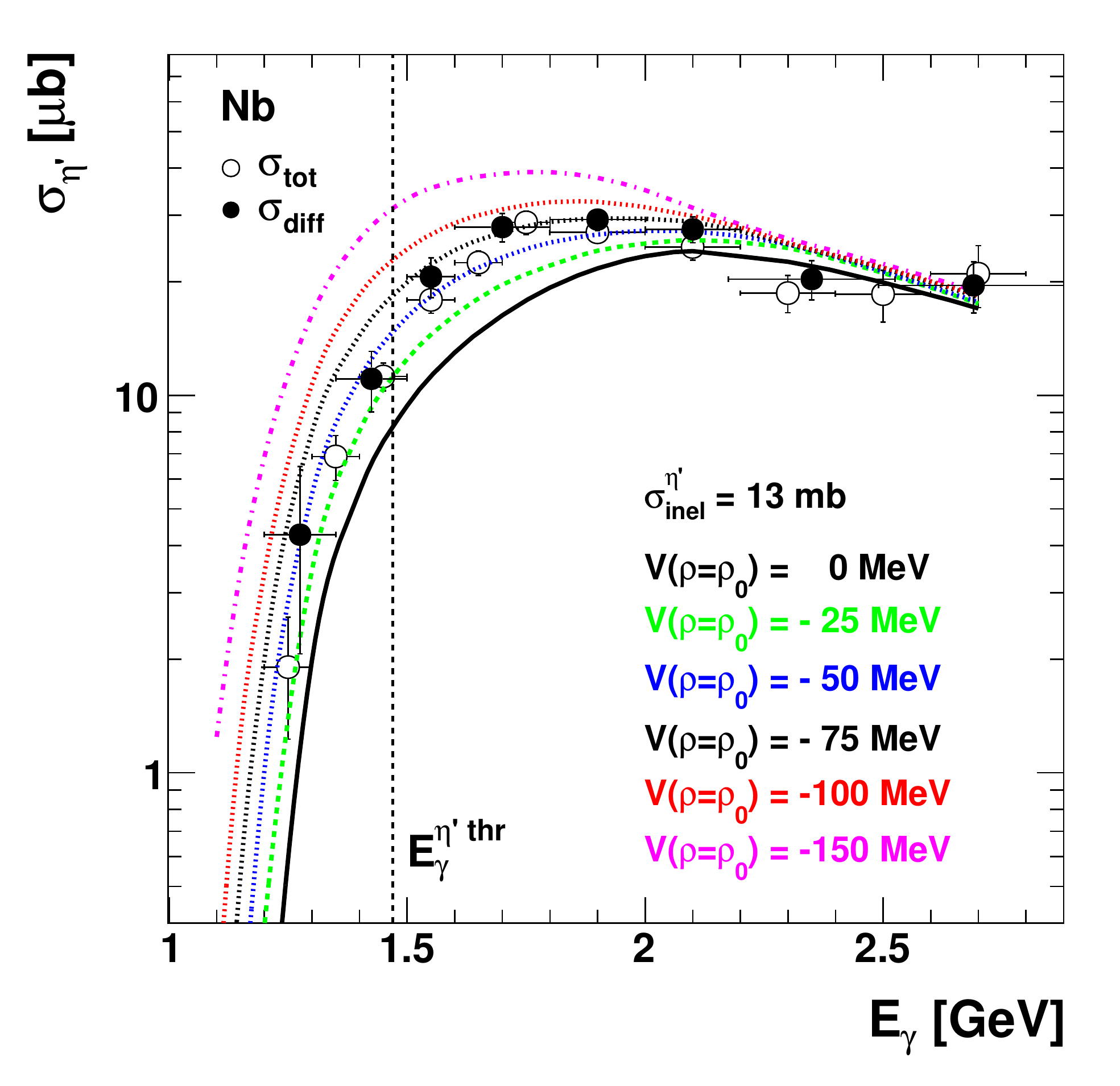}  \includegraphics{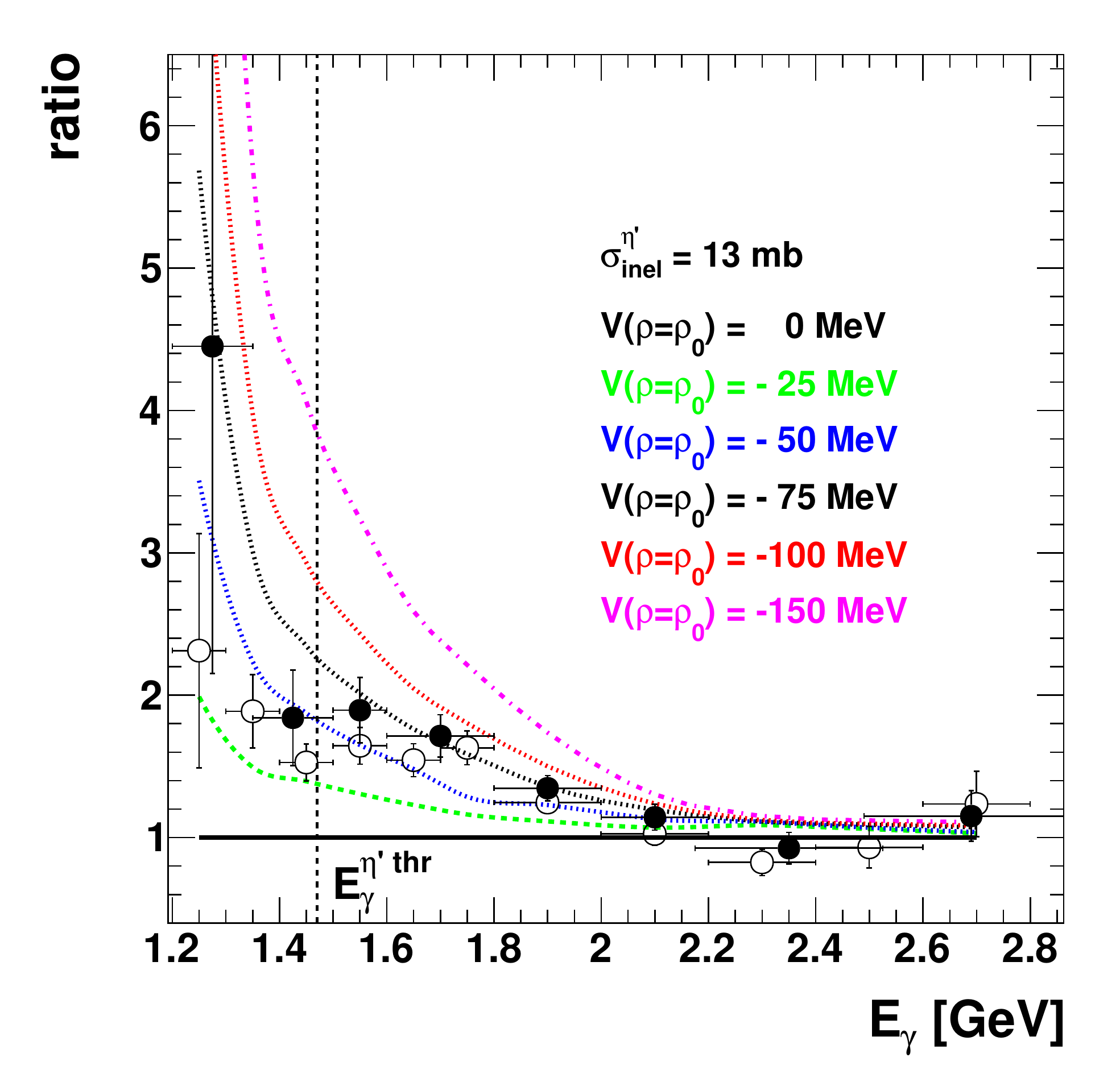}  \includegraphics[height=0.85\textheight]{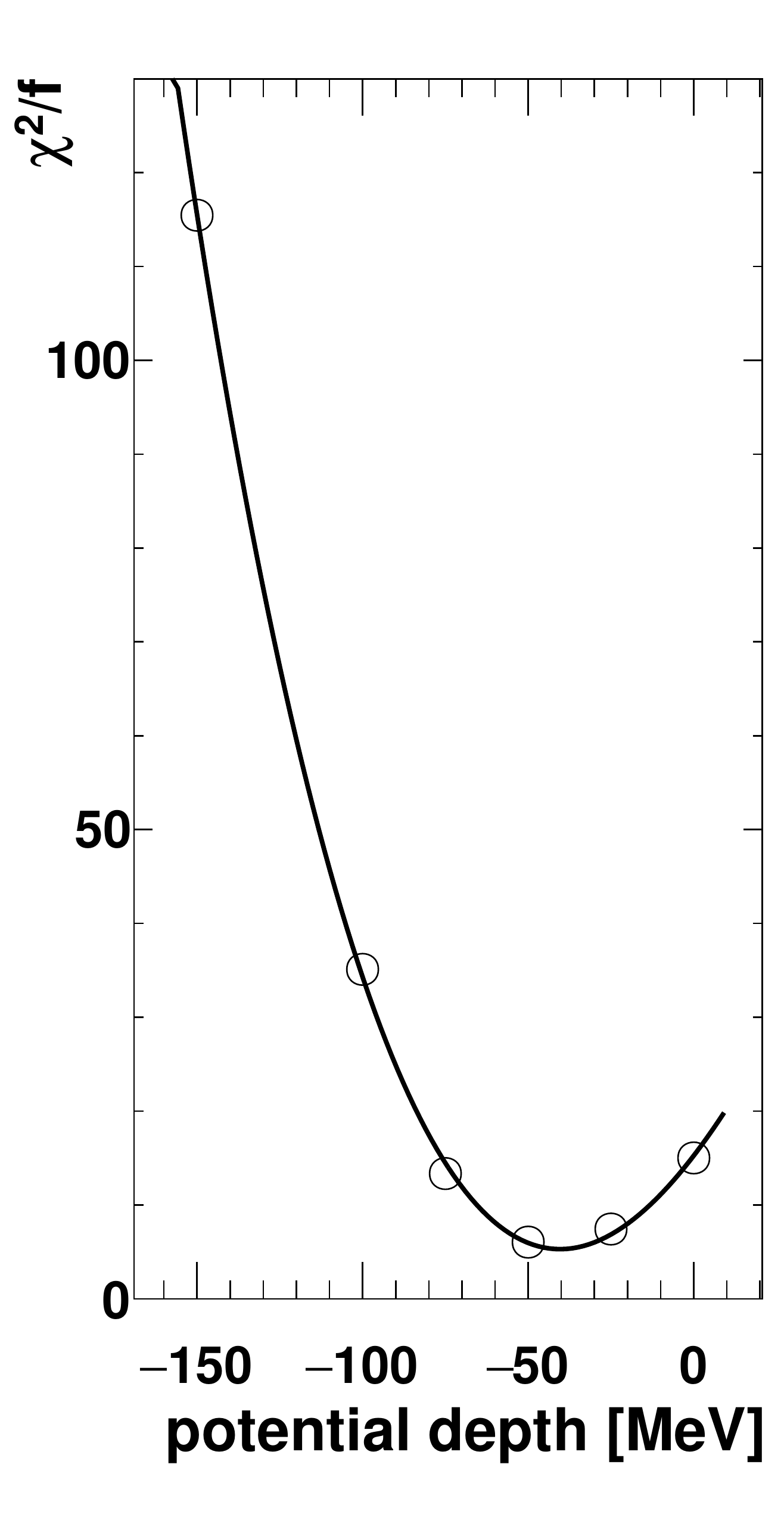} 
    }
\caption{(Color online) Left: Total cross section for $\eta^\prime$ photoproduction off Nb obtained by integrating the differential cross sections (full circles) and by direct measurement of the $\eta^\prime$ yield in different incident photon energy bins (open circles). The data are compared to calculations for an inelastic $\eta^\prime$-nucleon cross section
 $\sigma_{\text{inel}}^{\eta^\prime}$=13 mb \cite{Nanova_tr,Friedrich_ta} and for potential depths V=0 MeV (black line), -25 MeV (green), -50 MeV (blue), -75 MeV (black dashed), -100 MeV (red) and -150 MeV (magenta) at normal nuclear density, respectively, and using the full nucleon spectral function. All calculated cross sections have been multiplied by a factor 0.91 to match the experimentally observed cross section above E$_{\gamma} >$ 2.2 GeV (see text). This normalization is within the systematic error quoted for the experimental cross section. Middle: The experimental data and the predicted curves for V = - 25, -50, -75, -100 and -150 MeV divided by the calculation for the scenario V= 0 MeV and presented on a linear scale. Right: $\chi^2$ - fit of the data with the excitation functions calculated for the different scenarios over the full incident photon energy range.}
\label{fig:tot_Nb}
\end{center}
\end{figure*}

 The measured excitation function for photoproduction of $\eta^\prime$ mesons off Nb is compared in Fig.~\ref{fig:tot_Nb} (Left) to calculations within the first collision model \cite{Paryev}. These calculations are conceptually identical to the ones used for extracting the real part of the $\eta^\prime$-C potential \cite{Nanova_realC}. Using the measured differential cross sections for $\eta^\prime$ production off  the proton and neutron bound in deuterium \cite{Igal} as input, the cross section for $\eta^\prime$ photoproduction off Nb is calculated in an eikonal approximation, taking  the effect of the nuclear $\eta^\prime$ mean-field potential into account. While the cross section data go up to the highest incident photon energy of 2.9 GeV, the calculations do not extend beyond E$_{\gamma}$ = 2.7 GeV since the elementary $\eta^\prime $ photoproduction cross sections off the proton and neutron \cite{Igal} are only known up to this energy. The off-shell differential cross sections for the production of $\eta^\prime$ mesons with reduced in-medium mass off intranuclear protons and neutrons in the elementary reactions $\gamma \text{p} \rightarrow \eta^\prime$p and $\gamma \text{n} \rightarrow \eta^\prime$n are assumed to be given by the measured on-shell cross sections, using the reduced in-medium mass. The $\eta^\prime$ final-state absorption is taken into account by using a momentum independent, inelastic in-medium $\eta^\prime$N cross section of $\sigma_{\text{inel}}^{\eta^{\prime}}$=13$\pm$3 mb \cite{Friedrich_ta},  slightly larger but consistent within the errors with the result of previous transparency ratio measurements \cite{Nanova_tr}. The contribution of $\eta^\prime$ production from two-nucleon short-range correlations is implemented by using the total nucleon spectral function in the parametrisation by \cite{Efremov}. As in \cite{Nanova_realC}, the momentum-dependent optical potential from \cite{Rudy}, seen by the nucleons emerging from the nucleus in coincidence with the $\eta^\prime$ mesons, is accounted for. Furthermore, the Coulomb interaction of the outgoing proton and the residual nucleus is taken into account. The overall systematic uncertainties of the calculations are mainly given by the experimental input and the fits to the measured cross sections and are estimated to be of the order of 10-15\%.
 
\par
The calculations have been performed for six different scenarios assuming depths of the $\eta^\prime$ real potential at normal nuclear matter density of $V$ = 0, -25, -50, -75, -100\\and -150 MeV, respectively. The calculated cross sections have been scaled down - within the limits of the systematic uncertainties - by a factor of 0.91 to match the experimental excitation function data at incident photon energies above 2.2 GeV, where the difference between the various scenarios is very small. In the corresponding analysis of the C data \cite{Nanova_realC} a similar rescaling of the theoretical calculations had to be applied. We are not aware of any missing physics in the calculations which might explain this systematic difference between data and calculations. In view of the systematic errors of the cross section data (23$\%$) and the calculations (10-15$\%$) a discrepancy cannot be claimed. The highest sensitivity to the $\eta^\prime$ potential depth is found for incident photon energies near and below the production threshold on the free nucleon. As described in section \ref{sec:exp}, the photon flux has been enhanced below E$_{\gamma}$ = 1.5 GeV to achieve sufficient statistics in this particularly relevant energy regime where the cross sections are quite small. The excitation function data appear to be incompatible with $\eta^\prime$ mass shifts of -100 MeV and more at normal nuclear matter density, as more clearly seen in Fig.~\ref{fig:tot_Nb} (Middle), where the data and the calculations are shown on a linear scale after dividing by the curve corresponding to the scenario V= 0 MeV. A $\chi^{2}$-fit of the data (see Fig.~\ref{fig:tot_Nb} (Right)) over the full incident energy range with the excitation functions calculated for the different scenarios gives a potential depth of -(40$\pm$12) MeV. 

\subsection{Momentum distribution of the $\eta'$ mesons}
\label{sec:mom}
\begin{figure*}
\begin{center}
 \resizebox{1.0\textwidth}{!}
  {
   \includegraphics{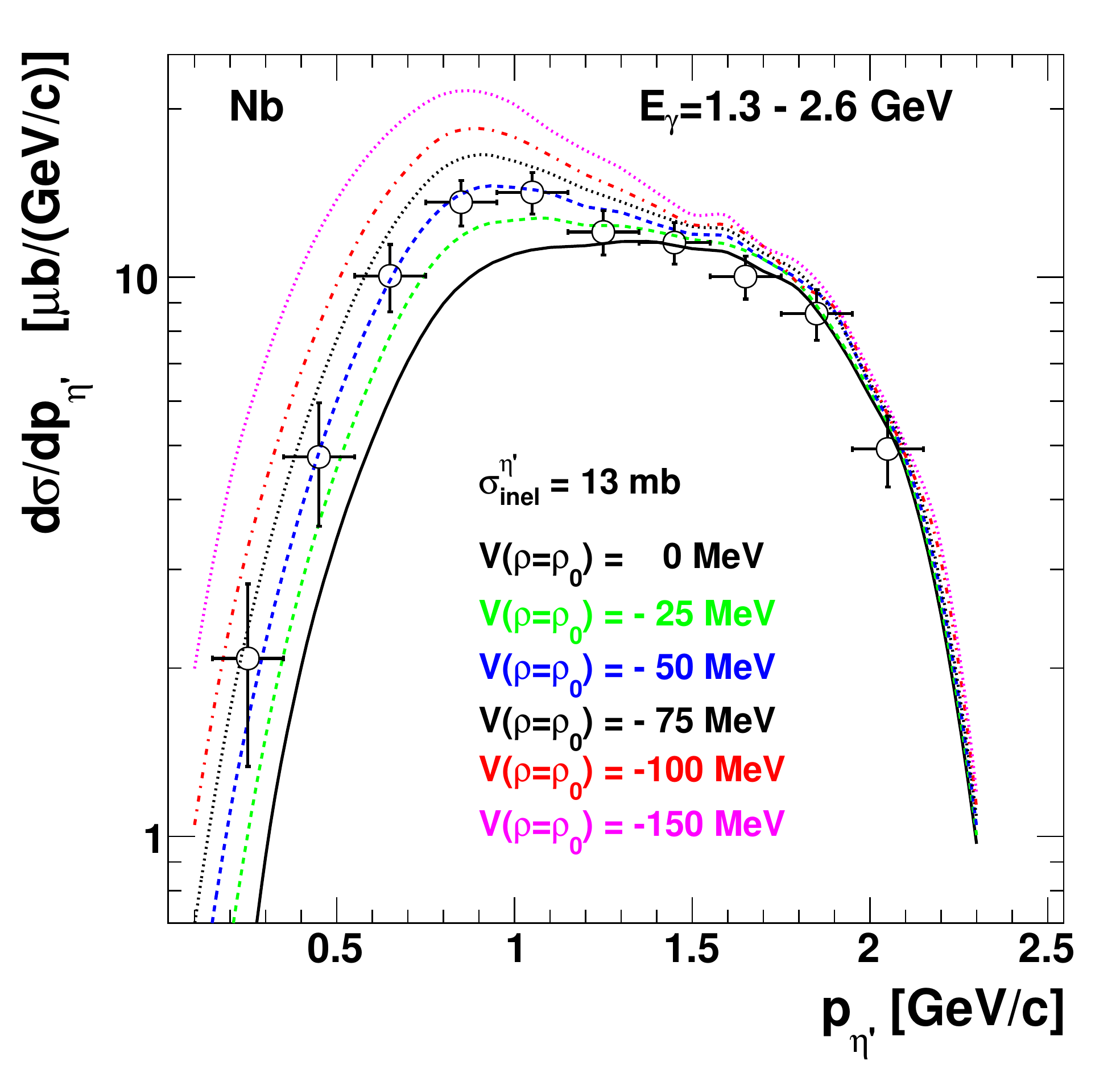}  \includegraphics{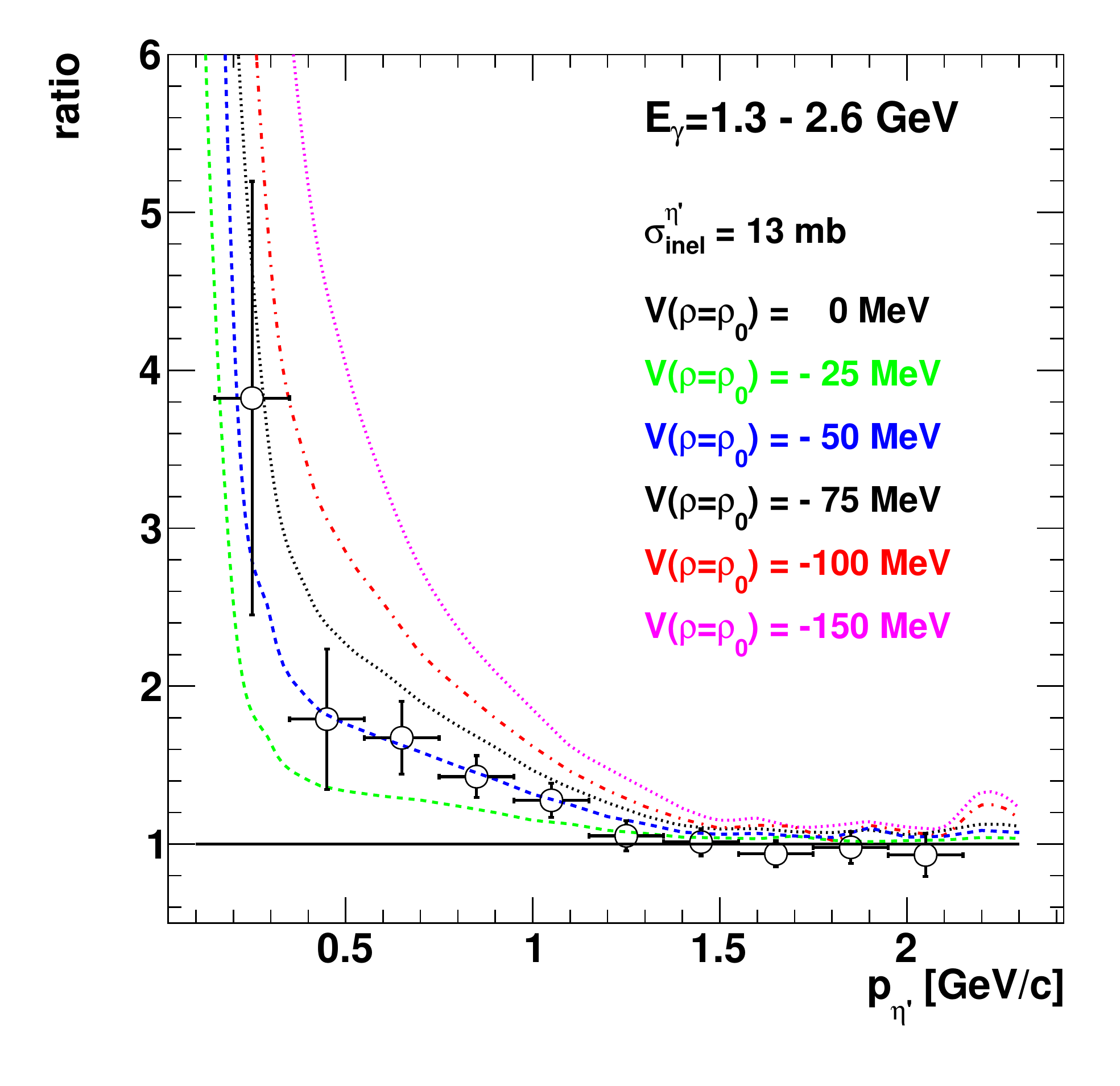}  \includegraphics[height=0.85\textheight]{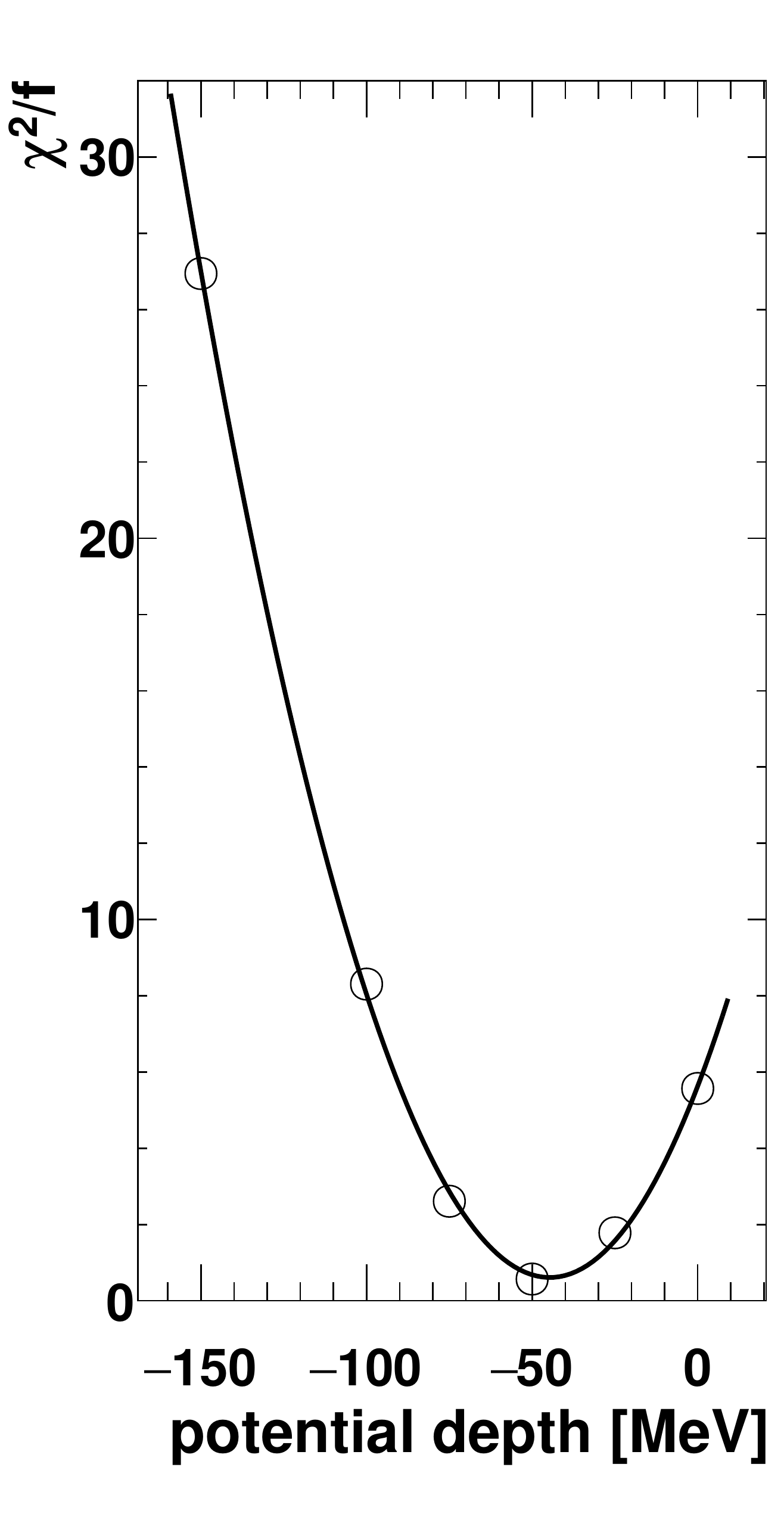} 
     }
\caption{(Color online) Left: Momentum distribution for $\eta^\prime$ photoproduction off Nb for the incident photon energy range 1.3-2.6 GeV. The calculations are for $\sigma_{\text{inel}}^{\eta^\prime}$=13 mb and for potential depths V = 0, -25, -50, -75, -100 and -150 MeV at normal nuclear density, respectively. All calculated cross sections have been multiplied by a factor 0.83 (see text). The color code is identical to the one in Fig.~\ref{fig:tot_Nb}. Middle:The experimental data and the predicted momentum distributions for V = - 25, -50, -75, -100 and -150 MeV divided by the calculation for the scenario V= 0 MeV and presented on a linear scale. Right: $\chi^2$ - fit of the data with the momentum distributions calculated for the different scenarios.  }
\label{fig:mom}
\end{center}
\end{figure*}

The measured momentum differential cross section for $\eta^\prime$ meson photoproduction off Nb is shown in Fig.~\ref {fig:mom} (Left). The average momentum is 1.14 GeV/$c$. Bin sizes of $\ge$ 0.2 GeV/$c$ have been chosen which are large compared to the momentum resolution of 25-50 MeV/$c$ deduced from the experimental energy resolution and from MC simulations. As described above, the momentum distribution of $\eta^\prime$ mesons is also sensitive to the $\eta^\prime$-potential depth. The $\eta^\prime$ momentum distributions have been calculated for the incident photon energy range 1.3-2.6 GeV and for different potential depths V = 0, -25, -50, -75, -100 and -150 MeV.  The comparison of these calculations with the data again seems to exclude strong $\eta^\prime$ mass shifts. In Fig.~\ref {fig:mom} (Middle) the experimental data and the predicted curves for V = - 25, -50, -75, -100 and -150 MeV are divided by the calculation for the scenario V= 0 MeV and presented on a linear scale. A $\chi^2$ - fit of the data (see Fig.~\ref{fig:mom} (Right)) with the momentum distributions calculated for the different scenarios gives an attractive potential of -(45$\pm$20) MeV.\\

\par
Combining the results from the analysis of the excitation function and the momentum distribution and by proper weighting of the errors a depth of the real part of the $\eta^\prime$-C and $\eta^\prime$-Nb optical potential of V$_0 (\rho = \rho_0) = -(37 \pm 10(stat)\pm10(syst))$ MeV and V$(\rho=\rho_0)$ = -(41 $\pm$10(stat)$\pm$15(syst)) MeV is obtained, respectively. The systematic error quoted is mainly due to uncertainties in normalizing the calculations to the data. The sensitivity of the result on this normalisation has been studied by varying the normalisation factor between 0.7 to 1.0 - well within the systematic errors of the cross section determinations.  This results for V$(\rho=\rho_0)$ are consistent with predictions of the $\eta^\prime$-nucleus potential depth within the Quark-Meson Coupling model (QMC) \cite{Bass} and with calculations in \cite{Nagahiro_Oset} but does not support larger mass shifts as discussed in \cite{Jido,Nagahiro,Kwon}.

\subsection{Comparison to $\eta^\prime $ photoproduction off carbon}
\begin{figure}
 \resizebox{0.5\textwidth}{!}
 {
   \includegraphics[height=0.9\textheight]{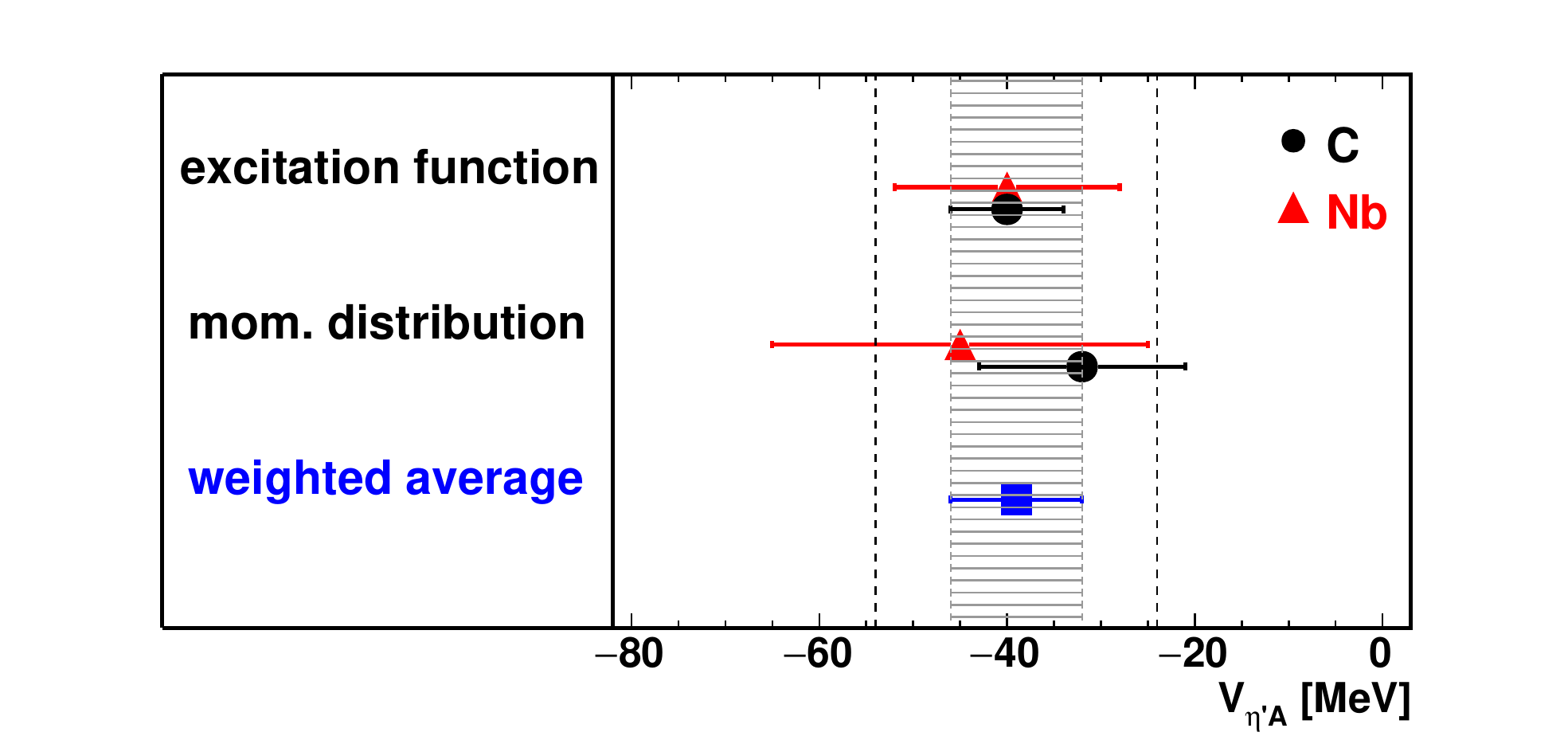} 
     }
\caption{Depths of the real part of the $\eta^\prime$-nucleus potential determined by analysing the excitation function and the momentum distributions for C \cite{Nanova_realC} (full black circles) and for Nb (this work, red triangles). The weighted overall average is indicated by a blue square and the shaded area. The vertical hatched lines mark the range of systematic uncertainties.}
\label{fig:comb_result}
\end{figure}

The depth of the potential determined in this work for the real part of the $\eta^\prime$-Nb interaction is compared in Fig.~\ref{fig:comb_result} to the result obtained for the 
$\eta^\prime$-C interaction \cite{Nanova_realC}. The values deduced by analysis of the excitation functions and the momentum distributions do agree for both nuclei within errors. Thus, the present result confirms the earlier observation from photoproduction of $\eta^\prime $ mesons off carbon that the mass of the $\eta^\prime $ meson is lowered by 
about 40 MeV in nuclei at saturation density, within the errors quoted for the potential depth. There is no evidence for a strong variation of the potential parameters with the nuclear mass number. Assuming that there is no mass number dependence the results separately obtained for both targets can be combined to the weighted average of V$(\rho=\rho_0)$= -(39$\pm$7(stat)$\pm$15(syst)) MeV, as shown in Fig.~\ref{fig:comb_result}. A simultaneous $\chi^2$ - fit to the 82 data points of both the C and Nb data sets, dominated by the C-data because of their better statistics, yields a potential depth of -(35 $\pm 15$) MeV. The modulus of the real part of the $\eta^\prime $ optical potential is larger than the modulus of the imaginary part of $\approx$ -10 MeV which still makes the $\eta^\prime $ meson a promising candidate for the search for mesic states. However, this search appears to be more complicated than previously assumed. Pronounced narrow structures in the excitation energy spectrum of the $\eta^\prime$ - nucleus system calculated for potential depths in the range of $\ge$ 100 MeV \cite{Nagahiro} are less likely to be expected in view of the present results.

\section{Conclusions}
From the analysis of the excitation function and momentum distribution of $\eta^\prime$ mesons in photoproduction off Nb the real part of the $\eta^\prime$-Nb optical potential has been determined. Within the model used, the present results are consistent with an attractive $\eta^\prime$-Nb potential with a depth of -(41$\pm$10(stat)$\pm$15(syst)) MeV under normal conditions ($\rho=\rho_{0}, T=0$). 
This result is consistent with an earlier determination of the $\eta^\prime$-C potential depth of -(37$\pm$10(stat)$\pm$10(syst)) MeV \cite{Nanova_realC} and confirms the (indirect) observation of a mass reduction of the $\eta^\prime$ meson in a strongly interacting environment at above conditions. The attractive $\eta^\prime$-nucleus potential may be strong enough to allow the formation of bound $\eta^\prime$-nucleus states. The search for such states is encouraged by the relatively small imaginary potential of the $\eta^\prime$~of $\approx$ -10 MeV~\cite{Nanova_tr}. Because of the relatively shallow $\eta^\prime$-nucleus potential found in this work, the search for $\eta^\prime$-mesic states may, however, turn out to be more difficult than initially anticipated on the basis of theoretical predictions. An experiment to search for $\eta^\prime$ bound states via missing mass spectroscopy~\cite{Kenta} has been performed at the Fragment Separator (FRS) at GSI and is being analyzed. A semi-exclusive measurement where observing the formation of the $\eta^\prime$-mesic state via missing mass spectroscopy is combined with the detection of its decay is ongoing at the LEPS2 facility (Spring8) \cite{Muramatsu} and is planned \cite{volker} at the BGO-OD setup \cite{Schmieden,Jude} at the ELSA accelerator in Bonn. A corresponding semi-exclusive experiment has also been proposed for the Super-FRS at FAIR~\cite{Nagahiro_Kenta}. The observation of $\eta^\prime$-nucleus bound states would provide further direct information on the $\eta^\prime$-nucleus interaction and the in-medium properties of the $\eta^\prime$ meson. 
 
\begin{acknowledgements}
We thank the scientific and technical staff at ELSA and the collaborating
institutions for their important contribution to the success of the
experiment. Detailed discussions with U. Mosel and J. Weil are acknowledged. This work was supported financially by the {\it
 Deutsche Forschungsgemeinschaft} within SFB/TR16 and by the {\it Schweize\-ri\-scher Nationalfonds}.
\end{acknowledgements}

\end{document}